\documentclass[aps,prl,letterpaper,10pt,amsmath,amssymb,superscriptaddress,nofootinbib,nolongbibliography,twocolumn]{revtex4-2}
\usepackage{graphicx}
\usepackage{xcolor}
\usepackage{mathrsfs}
\usepackage{braket}
\usepackage[colorlinks=true,linkcolor=blue,citecolor=blue,urlcolor=blue,bookmarks=true]{hyperref}
\usepackage[normalem]{ulem}
\usepackage{orcidlink}
\usepackage{diagbox}
\usepackage{CJKutf8}

\newcommand{\ad}{a^{\dagger}}
\newcommand{\tr}{\mathrm{Tr}}

\newcommand{\zh}[1]{\begin{CJK}{UTF8}{gbsn}#1\end{CJK}}

\begin{document}
\title{Superradiant phase transitions and a singular quantum critical endpoint in a cavity-coupled Heisenberg chain}
%\date{\today}

\author{Xiaotong Chen (\zh{陈晓桐})~\orcidlink{0009-0008-6656-872X}}
\affiliation{School of Physics Science and Engineering, Tongji University, Shanghai  200092, P. R. China}
\affiliation{Tsung-Dao Lee Institute,
Shanghai Jiao Tong University, Shanghai, 201210, China}

\author{Yiming Wang (\zh{王一鸣})~\orcidlink{0000-0002-7758-2921}}
\affiliation{Department of Physics \& Astronomy, Extreme Quantum Materials Alliance,
Smalley-Curl Institute, Rice University, Houston, Texas 77005, USA}

\author{Jianda Wu (\zh{吴建达})~\orcidlink{0000-0002-3571-3348}}
\altaffiliation{wujd@tongji.edu.cn}
\affiliation{School of Physics Science and Engineering, Tongji University, Shanghai  200092, P. R. China}

\begin{abstract}
Light–matter interaction in optical cavities has emerged as a promising platform for exploring and controlling quantum many-body phenomena. 
Yet understanding the coupling between cavity photons and strongly correlated quantum matter remains a theoretical challenge, particularly when quantum fluctuations play an essential role. Here, we investigate an antiferromagnetic spin-1/2 Heisenberg chain coupled to a single-mode cavity. We determine the phase diagram at zero and finite temperatures and uncover both continuous and first-order superradiant phase transitions. Remarkably, we find that the nonanalytic quantum fluctuations of the spin chain near the saturation field 
give rise to unconventional critical behavior, characterized by an exponent of 3/8 for 
the vanishing order-parameter jump, in contrast to the conventional exponent of 1/2 at finite temperatures. Furthermore, the competition between longitudinal and transverse spin fluctuations leads to a nonmonotonic temperature evolution of the phase boundaries. Our results demonstrate that quantum critical fluctuations in strongly correlated matter can fundamentally modify superradiant phase transitions and give rise to unconventional critical behavior in coupled light–matter systems.
\end{abstract}

\maketitle
\textit{Introduction}.\textemdash\
Coupling matter to a cavity provides a new route to engineer phases of both matter and light.
Various studies have demonstrated the potential of light-matter coupling for modifying the properties of matter and stabilizing experimentally elusive phases \cite{LiEckstein2020,Wei2025,Yang2026,Wang2026,Appugliese2022,Enkner2025,Jarc2023,Jiang2025,Cardoso2026}.
The interplay between photons and matter can also give rise to ordering of light.
One example is the superradiant (SR) phases of photons, in which a macroscopic occupation of photons is present \cite{Dicke1954,WangHioe1973,Sur2026}.
The competition and cooperation between SR order and matter order have the ability to engineer orders in strongly correlated systems as well as enhance squeezing and light-matter entanglement near criticality \cite{Zwettler2025,Sur2026}.
It remains, however, an open question how the quantum critical behavior of many-body systems shapes the phase boundaries and the corresponding critical behavior of the light-matter coupled system.

The antiferromagnetic spin-$1/2$ Heisenberg (XXX) chain provides a controllable setting for this question.
Its gapless ground state exhibits power-law decaying spin correlations, 
while a uniform magnetic field drives it 
toward a fully polarized state at the saturation field $h_s=2J$~\cite{Bethe1931,Griffiths1964,YangYang1966}.
The Bethe ansatz and thermodynamic Bethe ansatz (TBA) provide exact equations for its ground-state energy and 
finite-temperature ($T$)  thermodynamics, respectively~\cite{Takahashi1971,Giamarchi2004}.
Spin-rotation invariance combines the static and cavity-induced fields into a single effective field, making it possible to trace how the field-induced quantum
phase transition and its thermal fluctuations shape the superradiant phase boundaries.

In this letter, we combine the chain's thermodynamics with a static saddle-point treatment of the cavity field to determine the zero- and finite-$T$ phase diagrams.
At zero $T$, an explicit coexistence relation connects the first-order boundary to the gapless ground state energy 
of the XXX chain near full polarization and identifies the origin and path dependence of the singular endpoint scaling.
At finite $T$s, we use a local $\phi^6$ free energy to describe how the first-order and continuous boundaries meet, 
which is attributed to the competition of spin fluctuations of the XXX chain in different directions.
A comparison of the fixed-$T$ phase diagrams 
reveals the non-monotonic endpoint drift resulted by 
the thermal evolution between balance of the competitive fluctuations.

\textit{The model}.\textemdash\
We consider the XXX chain coupled to a single-mode cavity and subjected to a static magnetic field perpendicular to the cavity-induced field,
\begin{equation}
 H=\omega\ad a+J\sum_j\vec S_j\cdot\vec S_{j+1}
 -\frac{g(a+\ad)}{\sqrt N}\sum_j S_j^z-h\sum_j S_j^x,
 \label{eq:model}
\end{equation}
with number of spins $N$, the cavity frequency $\omega>0$, the light--matter coupling strength $g\geq0$, the static field $h\geq0$, and the antiferromagnetic coupling $J>0$.
For convenience, we set $J=\hbar=k_B=1$.
The factor $N^{-1/2}$ keeps the light--matter coupling energy extensive when the photon occupation is proportional to $N$.
For $J=0$, Eq.~\eqref{eq:model} reduces to the Dicke model of noninteracting spins coupled to a single-mode cavity.
We write $H_\mathrm{s}(h)=J\sum_j\vec S_j\cdot\vec S_{j+1}-h\sum_j S_j^x$ for the uncoupled spin Hamiltonian, whose ground state is gapless for $0\leq h\le 2J$ and fully polarized for $h\ge2J$, with the saturation transition at $h=2J$.

Within the static saddle-point approximation for the cavity field, the equilibrium free-energy density $ f_{\mathrm{eq}}(T,h,g)$ is obtained from (see End Matter)
\begin{align}
 f_{\mathrm{eq}}(T,h,g)&=\min_{\phi\in\mathbb R}\mathcal F(\phi,T,h,g),\\
 \mathcal F(\phi,T,h,g)&=\frac{\omega\phi^2}{4g^2}+f_{\mathrm{lm}}(T,\phi,h),
 \label{eq:freeenergy}
\end{align}
where $f_{\mathrm{lm}}(T,\phi,h)$ is the free energy density for the Hamiltonian $H_{\mathrm{lm}}=H_\mathrm{s}(h)-\phi\sum_jS_j^z$.
The minimizing field $\phi_0$ is the superradiant (SR) order parameter, and in a symmetry-broken equilibrium state $\phi_0=g\langle a+\ad\rangle/\sqrt N$
where $\braket{\cdot}$ denotes the thermal average and reduces to ground state expectation value at zero $T$.
The photon density at equilibrium follows $n_{\mathrm{ph}}\equiv\langle\ad a\rangle/N=\phi_0^2/(4g^2)$.
The combined transformation $a\mapsto-a$ and a spin rotation by $\pi$ about the $x$ axis is a $\mathbb Z_2$ symmetry, so the SR phase has symmetry-related minima $\pm\phi_0\neq0$ and a macroscopic photon occupation, whereas the normal phase has $\phi_0=0$ and a vanishing photon density in the thermodynamic limit.
Spin-rotation invariance of $H_s$ implies $f_\mathrm{lm}(T,\phi,h)=f_{\mathrm{lm}}(T,\sqrt{h^2+\phi^2})$,
where $f_{\mathrm{s}}$ can be obtained from the Bethe ansatz at $T=0$ and from the TBA at $T>0$.

\textit{Zero-$T$ phase diagram}.\textemdash\
\begin{figure}[t]
 \centering
 \includegraphics[width=0.96\linewidth]{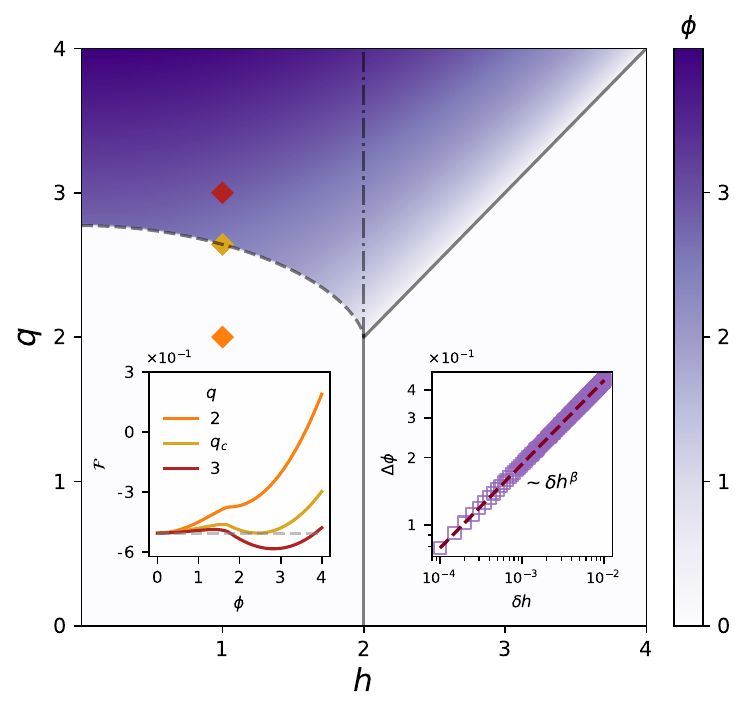}
 \caption{Zero-$T$ phase diagram. The normal phase lies below the superradiant boundary. The dashed boundary for $h<2$ is first order, while the solid boundary for $h>2$ is continuous, with $q =  h$. The vertical segment at $h=2$ separates the gapless and fully polarized normal regimes. The three boundaries meet at $(h,q)=(2,2)$. 
 The dash-dot line in the SR region separates two distinct regimes
 for the critical behavior of $\phi_0$: path-dependent scaling with continuous change of exponent on the left and universal scaling on the right.
 The left inset shows the free-energy-density functional landscape for the $h=1$ case, where $q_c \approx 2.642$.
 The corresponding points are marked in the phase diagram by markers in matching colors.
 The right inset shows the jump $\Delta\phi$ versus $\delta h=2-h$ along dashed boundary. The fitted exponent $\beta \approx 0.376$ is consistent with analytical result $3/8$ [Eq.~(\ref{eq:zeroT_jump_scaling})].}
 \label{fig:zeroT_phase_diagram}
\end{figure}
At zero $T$, $f_{\mathrm{lm}}$ reduces to the ground-state energy density $e_{\mathrm{lm}}$, and we determine the phase diagram in Fig.~\ref{fig:zeroT_phase_diagram} by minimizing $\mathcal F$.

For $h\geq2$, the effective field satisfies $\sqrt{h^2+\phi^2}\geq2$ for every $\phi$, and the free-energy-density functional reads
\begin{equation}
 \mathcal F=\frac{\phi^2}{4q}
 +\frac14-\frac{\sqrt{h^2+\phi^2}}{2},
 \label{eq:saturated}
\end{equation}
where $q\equiv g^2/\omega$.
Minimizing $\mathcal{F}$ gives
\begin{equation}
 |\phi_0|=\begin{cases}
 0,&q\leq h,\\
 \sqrt{q^2-h^2},&q>h.
 \end{cases}
 \label{eq:continuous}
\end{equation}
The phase boundary is therefore $q_c=h$.
At fixed $h\geq2$, the order parameter grows continuously as $|\phi_0|\propto(q-q_c)^{1/2}$ on entering the SR phase, as expected for a mean-field continuous transition because the free-energy-density functional has the same form as in the Dicke model~\cite{WangHioe1973}.

For $h<2$, the normal state is gapless, and minimization instead gives a first-order 
SRPT into a state fully polarized along the total field direction [Fig.~\ref{fig:zeroT_phase_diagram}].
The cavity-induced field raises $\sqrt{h^2+\phi^2}$ beyond saturation and lowers 
$\mathcal{F}$, whereas the accompanying macroscopic photon occupation incurs a cavity-energy cost that raises $\mathcal{F}$.
The competition between them determines the global minimum configuration [Fig.~\ref{fig:zeroT_phase_diagram}].
At the phase boundary the energy gain balances this cost, and the gapless normal and polarized SR states have equal free energy densities.
This coexistence gives
\begin{equation}
 e_{\mathrm{s}}(h) =\frac{\Delta\phi^2}{4q_c}
 +\frac14-\frac{\sqrt{h^2+\Delta\phi^2}}{2},
 \; \Delta\phi=\sqrt{q_c^2-h^2},
 \label{eq:coexistence}
\end{equation}
where $q_c(h)=g_c^2(h)/\omega$, and $\Delta\phi$ labels the first order jump of the photon condensation when
the system arrives at the transition boundary.

At $(h,q)=(2,2)$, the first-order SRPT boundary meets the continuous transition boundary and the normal-state saturation line.
We refer to this termination of the first-order SRPT boundary as the singular zero-$T$ endpoint.
For small $\delta h=2-h>0$, the energy density of the gapless ground state for $H_\mathrm{s}$ follows~\cite{Griffiths1964} (see End Matter)
\begin{equation}
e_{\mathrm{s}}(h) \sim -3/4 + \delta h/2 - A (\delta h)^{3/2},
 \label{eq:saturation_deficit}
\end{equation}
where $A = 2\sqrt2/(3\pi)$.
Eq.~\eqref{eq:coexistence} then gives
\begin{equation}
(q_c-2)\propto\delta h^{3/4}\; \text{and} \; \Delta\phi\propto\delta h^{3/8}.
 \label{eq:zeroT_jump_scaling}
\end{equation}
The scaling exponent changes when the path to approach the endpoint deforms.
To make this dependence explicit, we consider a path inside the SR phase with $\delta q\equiv q-2=B\delta h^p$, where $B>0$ and $0<p<3/4$.
The polarized SR minimum satisfies
\begin{equation}
 |\phi_0|^2=(\delta q+\delta h)(4+\delta q-\delta h),
 \label{eq:zeroT_path}
\end{equation}
so $|\phi_0|\simeq2\sqrt{\delta q}\propto\delta h^{p/2}$, which is distinct from the exponent $3/8$ obtained along the first-order boundary.
When $h\geq2$, approaching to the endpoint inside the SR phase along a general path $q=\gamma(h)$ gives that $|\phi_0|^2=\gamma(h)^2-h^2$.
The constraints imposed on $\gamma(h)$ are that $\gamma(2)=2$ and $\gamma^{\prime}(2)\geq1$.
Therefore, up to the leading order of $h-2$, we have $|\phi_0|\propto(h-2)^{1/2}$. 
This scaling exponent does not depend on the the approach path to the endpoint.

\textit{Finite-$T$ phase diagram}.\textemdash\
\begin{figure*}[t]
 \centering
 \includegraphics[width=0.95\textwidth]{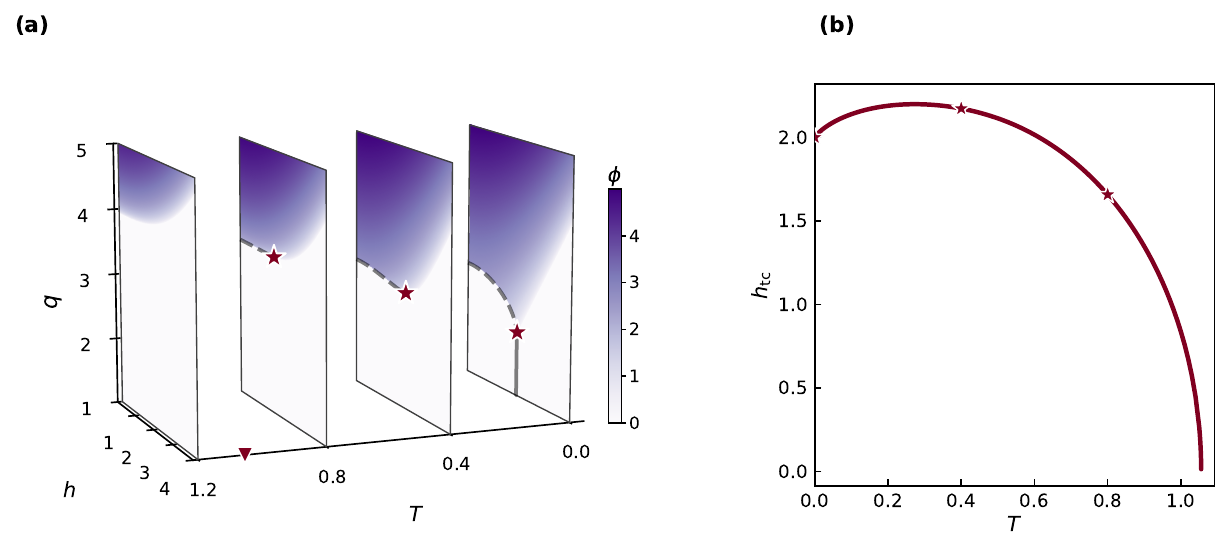}
 \caption{(a) Phase diagrams at finite $T$, including $T=0$ for comparison. Stars mark the meeting of first-order
 and continuous superradiant boundaries. 
 At $T=1.2$, the phase transition is continuous.
 The triangle marker in the $T$-axis denotes the temperature above which the first-order phase transition no longer occurs.
 The normal-state saturation transition line at $T=0$ becomes a smooth crossover at finite $T$. 
 (b) Endpoint locations obtained from (a), showing their non-monotonic drift with $T$.}
 \label{fig:finite_temperature_pt}
\end{figure*}
At finite $T$, the saturation transition of the isolated chain becomes a smooth crossover, 
since the free energy and all its field derivatives are analytic.
The cavity-coupled system nevertheless retains a sharp SRPT in the thermodynamic limit.
We first fix $T$ and determine the corresponding phase diagram in the $(h,q)$ plane by minimizing Eq.~\eqref{eq:freeenergy}.
In a fixed-$T$ slices containing transitions of both orders, 
the first-order transition boundary meets the continuous transition boundary at endpoint $(h_{\mathrm{tc}}(T),q_{\mathrm{tc}}(T))$, as illustrated in [Fig.~\ref{fig:finite_temperature_pt}(a)].
The subscript ``tc'' indicates the fact that the endpoints are tricritical points, as we will show later. 
As $h$ approaches $h_{\mathrm{tc}}$ along this boundary at fixed $T$, the first-order jump vanishes.
Near such an endpoint, expanding $\mathcal{F}(\phi)$ gives
\begin{equation}
 \mathcal F(\phi)-\mathcal F(0)
 =\frac r2\phi^2+\frac u4\phi^4+\frac v6\phi^6+O(\phi^8).
 \label{eq:thermal_landau_main}
\end{equation}
Here $r=r(T,h,q)$, whereas $u=u(T,h)$ and $v=v(T,h)$ depend only on the chain (see End Matter).
We first discuss the free-energy landscape described by this $\phi^6$ theory, assuming $v>0$ near the endpoint.

The quadratic coefficient $r$ measures the local stability of the normal state.
For $u>0$, the minimum remains at $\phi=0$ while $r>0$ and moves continuously to nonzero $\phi$ when $r$ becomes negative.
The continuous transition therefore occurs at $r=0$, provided a distant minimum of the full free energy does not preempt it.
For $u<0$, the negative quartic term favors finite $|\phi|$, while the positive sixth-order term stabilizes the SR minima, which appear as $r$ decreases below $u^2/(4v)$.
Meanwhile, $\phi=0$ remains locally stable for $r>0$.
The first-order phase transition happens when these two local minima become degenerate. 
This finite-$T$ coexistence implies (see End Matter)
\begin{equation}
 \Delta\phi^2=-\frac{3u}{4v},\qquad
 r=\frac{3u^2}{16v}.
 \label{eq:finiteT_coexistence}
\end{equation}
Thus the first-order transition occurs before the normal state loses local stability, and the minima $0$ and $\pm\Delta\phi$ are separated by barriers of height $(-u)^3/(96v^2)$ to leading order (see End Matter).
Since the jump and the barrier vanish together as $u$ approaches $0$, the meeting of the first-order and continuous boundaries has $r=u=0$ with $v>0$ and is a tricritical point within the static saddle-point theory.
We use the term thermal endpoint to refer to this termination of the first-order SRPT boundary at fixed $T$.

The coefficients of the $\phi^6$ theory have a direct magnetic interpretation. 
When $h>0$ (see End Matter),
\begin{equation}
 r=\frac{1}{2q}-\chi_\perp,\qquad
 u=\frac{\chi_\perp-\chi_\parallel}{2h^2},
 \label{eq:thermal_response_main}
\end{equation}
where $m(T,h)=-\partial_h f_{\mathrm{s}}(T,h)$, $\chi_\parallel=\partial_hm$, and $\chi_\perp=m/h$.
The $\chi_\parallel$ measures changes in the magnetization magnitude, 
whereas the $\chi_\perp$ characterizes the response to a small field perpendicular to the static field.
On the first-order SRPT side, the competition between energy cost of macroscopic photon occupation and energy lowering from the transverse response maintains the normal state as local minimum, while $\chi_\parallel>\chi_\perp$ leads to a global SR minimum. 
This balance reverses on the continuous side. 
At fixed $T$, a nonzero-field endpoint is found from ($v> 0$ is required for stability)
\begin{equation}
 \chi_\parallel(T,h_{\mathrm{tc}})=\chi_\perp(T,h_{\mathrm{tc}}),\ q_{\mathrm{tc}}=\frac{1}{2\chi_\perp(T,h_{\mathrm{tc}})}.
 \label{eq:thermal_endpoint}
\end{equation}
The expression for $v$ and the vanishing static field limit are given in the End Matter.
We locate these thermal endpoints by solving Eq.~\eqref{eq:thermal_endpoint}.
At moderate and higher $T$s, finite-size exact diagonalization (ED) evaluates the magnetization and susceptibility and efficiently resolves the sign change of $u$
[In this regime, resolving the same sign change with string-truncated TBA can be costly even when the free energy appears converged].
At low $T$s near saturation, we instead use TBA to avoid appreciable ED finite-size effects.
The numerical details are described in the End Matter.

Thermal fluctuations modify the scaling behavior of the order parameter as the endpoints are approached. 
For a generic endpoint at $h_{\mathrm{tc}}>0$, we write $\delta h_T=h_{\mathrm{tc}}(T)-h>0$ and $u=-u_h\delta h_T+O(\delta h_T^2)$, where $u_h=(\partial_hu)|_{h=h_{\mathrm{tc}}}$ (see End Matter).
Eq.~\eqref{eq:finiteT_coexistence} therefore gives
\begin{equation}
 \Delta\phi\propto\delta h_T^{1/2}.
 \label{eq:thermal_jump_scaling}
\end{equation}
This exponent differs from the zero-$T$ value of $3/8$.
As at zero $T$, the measured power law depends on the approach to the endpoint.
At fixed $T$ and $h=h_{\mathrm{tc}}$, increasing $q$ beyond $q_{\mathrm{tc}}$ keeps $u=0$ and decreases $r$ below zero, giving $|\phi_0|=(-r/v)^{1/4}$ to leading order.
This exponent describes the continuous growth of the order parameter within the SR phase, 
whereas $1/2$ characterizes how the first-order jump vanishes along the coexistence line.
The endpoint at $h_{\mathrm{tc}}=0$ is described by the analytic $\phi^6$ theory and the same conditions $r=u=0$ with $v>0$.
Using the regular zero-field coefficients (see End Matter), 
its temperature $T_{\mathrm{tc}}^{(0)} \approx 1.056$ is determined by $u(T_{\mathrm{tc}}^{(0)},0)=0$, 
and its coupling $q_{\mathrm{tc}}^{(0)}$ follows from $r(T_{\mathrm{tc}}^{(0)},0,q_{\mathrm{tc}}^{(0)})=0$.

We now compare the fixed-$T$ diagrams.
The endpoint field $h_{\mathrm{tc}}(T)$ initially moves above $2$, then turns back and moves below $2$ as temperature increases [Fig.~\ref{fig:finite_temperature_pt}(b)].
Heating therefore initially extends the range of fields supporting a first-order SRPT before suppressing it at higher $T$s.
At $T=1.2$, the boundary shown is continuous throughout.
This non-monotonic evolution follows from the response compeition in Eq.~\eqref{eq:thermal_endpoint}.
At low $T$, the large longitudinal response $\chi_\parallel$ near saturation 
extends above $h=2$ due to the thermally populated excitations.
Further above saturation, $\chi_\parallel$ falls rapidly, while $\chi_\perp=m/h$ remains finite, placing their crossing on the high-field side.
At higher $T$, thermal broadening weakens the saturation feature and changes the magnetization over a broader field range. 
The endpoint data show that the crossing moves to lower fields as the region with $\chi_\parallel>\chi_\perp$ contracts.
In the high-$T$ limit, the leading magnetization is that of independent thermal spins, $m\simeq\tfrac12\tanh[h/(2T)]$.
This expression gives $\chi_\parallel<\chi_\perp$ for every $h>0$, consistent with continuous superradiant transitions in this limit.

\textit{Discussion and conclusions}.\textemdash\ 
At zero $T$, near the saturation field, long-range spin fluctuations compete with those induced by light-matter coupling, producing a singularity at the endpoint.
Approaching the endpoint along different paths yields different scaling from the first-order side, whereas the scaling is universal from the continuous side.
Because of this singularity, the endpoint is not an ordinary analytic tricritical point, unlike its finite-$T$ counterpart.
Even though the expansion at exactly $h=2$ has a finite positive quartic coefficient (see End Matter), it fails as soon as $h<2$.

At finite $T$, the rounding of the saturation crossover admits an analytic free energy near the endpoint, and generic endpoints have the $\phi^6$ tricritical structure.
Along the first-order phase transition boundary at fixed $T$, the order-parameter jump vanishes with exponent $1/2$ on approaching a generic endpoint.
Comparing different $T$s reveals a second consequence of thermal fluctuations: the enhanced $\chi_\parallel$ 
first extends beyond the zero-$T$ saturation field and then weakens as thermal effects grow. Its changing balance with $\chi_\perp$ accounts for the observed non-monotonic endpoint drift.
Thus, thermal fluctuations not only alter the scaling behavior of the order parameter near the endpoints but also shift the endpoints themselves.

Although the scaling behavior of the order parameter near the endpoint is sensitive to thermal fluctuations, the $3/8$ exponent can be captured when spin fluctuations dominate.
In the regime $T\ll\delta h\ll1$ with $\delta h=2-h$, the relative thermal correction to the normal-state free-energy density is $O[(T/\delta h)^2]$ and thermal excitations of the coexisting SR state are exponentially suppressed (see End Matter).
Consequently, along the first-order phase transition boundary,
\begin{equation}
 \Delta\phi(T,h)=\Delta\phi(0,h)
 \left[1+O\!\left(\frac{T^2}{\delta h^2}\right)\right],
 \label{eq:lowT_jump_recovery}
\end{equation}
where $\Delta\phi(0,h)\propto\delta h^{3/8}$ near saturation.
Thus the zero-$T$ exponent can be extracted from finite-$T$ coexistence curves with $\delta h\to0$ and $T/\delta h\to0$.
When $T$ becomes comparable to $\delta h$, the low-$T$ expansion does not apply.
Continuing along the coexistence line toward the shifted endpoint at fixed $T>0$ ultimately gives a scaling behavior with exponent $1/2$.
These limits explain the sudden change of the endpoint scaling from zero to finite $T$.
Thermal fluctuations shift the endpoint away from the region where the long-range quantum fluctuations dominate.

Quantum materials already provide experimental platforms.
One-dimensional spin-$1/2$ XXX models have been identified in materials such as copper pyrazine dinitrate ($\mathrm{CuPzN}$) and $\mathrm{KCuF_3}$~\cite{Hammar1999,Stone2003,Kono2015,Breunig2017,Lake2013}.
The field-induced saturation transition is realized in $\mathrm{CuPzN}$, making it a potential candidate for experimentally exploring the intertwined phase transitions induced by light-matter coupling.

To summarize, we study the equilibrium phase diagram of the Heisenberg model coupled to a single-mode cavity at both zero and finite $T$s.
Our results demonstrate that the competition between the long-range quantum fluctuations near the critical point 
and the long-range fluctuations induced by the cavity determines the overall structure of 
the phase diagram and gives rise to novel critical behaviors. 
Thermal fluctuations drive the endpoint away from the critical point, allowing a $\phi^6$ description. 
Nevertheless, within an appropriate temperature window, the consequences of the singular critical point can still be captured.
\par
\textit{Acknowledgment}\textemdash\
This work is sponsored by the National Natural Science Foundation of China Nos. 12274288, 12450004, 
the Innovation Program for Quantum Science and Technology Grant No. 2021ZD0301900,
and the Fundamental Research Funds for the Central
Universities. 
J.W. acknowledges the hospitality of Wilczek Quantum Center at Shanghai Institute for Advanced Studies of University of Science and Technology of China.

\bibliography{cavity}

\begin{thebibliography}{27}%
\makeatletter
\providecommand \@ifxundefined [1]{%
 \@ifx{#1\undefined}
}%
\providecommand \@ifnum [1]{%
 \ifnum #1\expandafter \@firstoftwo
 \else \expandafter \@secondoftwo
 \fi
}%
\providecommand \@ifx [1]{%
 \ifx #1\expandafter \@firstoftwo
 \else \expandafter \@secondoftwo
 \fi
}%
\providecommand \natexlab [1]{#1}%
\providecommand \enquote  [1]{``#1''}%
\providecommand \bibnamefont  [1]{#1}%
\providecommand \bibfnamefont [1]{#1}%
\providecommand \citenamefont [1]{#1}%
\providecommand \href@noop [0]{\@secondoftwo}%
\providecommand \href [0]{\begingroup \@sanitize@url \@href}%
\providecommand \@href[1]{\@@startlink{#1}\@@href}%
\providecommand \@@href[1]{\endgroup#1\@@endlink}%
\providecommand \@sanitize@url [0]{\catcode `\\12\catcode `\$12\catcode
  `\&12\catcode `\#12\catcode `\^12\catcode `\_12\catcode `\%12\relax}%
\providecommand \@@startlink[1]{}%
\providecommand \@@endlink[0]{}%
\providecommand \url  [0]{\begingroup\@sanitize@url \@url }%
\providecommand \@url [1]{\endgroup\@href {#1}{\urlprefix }}%
\providecommand \urlprefix  [0]{URL }%
\providecommand \Eprint [0]{\href }%
\providecommand \doibase [0]{https://doi.org/}%
\providecommand \selectlanguage [0]{\@gobble}%
\providecommand \bibinfo  [0]{\@secondoftwo}%
\providecommand \bibfield  [0]{\@secondoftwo}%
\providecommand \translation [1]{[#1]}%
\providecommand \BibitemOpen [0]{}%
\providecommand \bibitemStop [0]{}%
\providecommand \bibitemNoStop [0]{.\EOS\space}%
\providecommand \EOS [0]{\spacefactor3000\relax}%
\providecommand \BibitemShut  [1]{\csname bibitem#1\endcsname}%
\let\auto@bib@innerbib\@empty
%</preamble>
\bibitem [{\citenamefont {Li}\ and\ \citenamefont
  {Eckstein}(2020)}]{LiEckstein2020}%
  \BibitemOpen
  \bibfield  {author} {\bibinfo {author} {\bibfnamefont {J.}~\bibnamefont
  {Li}}\ and\ \bibinfo {author} {\bibfnamefont {M.}~\bibnamefont {Eckstein}},\
  }\href {https://doi.org/10.1103/PhysRevLett.125.217402} {\bibfield  {journal}
  {\bibinfo  {journal} {Phys. Rev. Lett.}\ }\textbf {\bibinfo {volume} {125}},\
  \bibinfo {pages} {217402} (\bibinfo {year} {2020})}\BibitemShut {NoStop}%
\bibitem [{\citenamefont {Wei}\ \emph {et~al.}(2025)\citenamefont {Wei},
  \citenamefont {Yang},\ and\ \citenamefont {Jiang}}]{Wei2025}%
  \BibitemOpen
  \bibfield  {author} {\bibinfo {author} {\bibfnamefont {C.}~\bibnamefont
  {Wei}}, \bibinfo {author} {\bibfnamefont {L.}~\bibnamefont {Yang}},\ and\
  \bibinfo {author} {\bibfnamefont {Q.-D.}\ \bibnamefont {Jiang}},\ }\href
  {https://doi.org/10.1103/8qx2-xxh2} {\bibfield  {journal} {\bibinfo
  {journal} {Phys. Rev. Lett.}\ }\textbf {\bibinfo {volume} {135}},\ \bibinfo
  {pages} {236901} (\bibinfo {year} {2025})}\BibitemShut {NoStop}%
\bibitem [{\citenamefont {Yang}\ \emph {et~al.}(2026)\citenamefont {Yang},
  \citenamefont {Cardoso}, \citenamefont {Hansson},\ and\ \citenamefont
  {Jiang}}]{Yang2026}%
  \BibitemOpen
  \bibfield  {author} {\bibinfo {author} {\bibfnamefont {L.}~\bibnamefont
  {Yang}}, \bibinfo {author} {\bibfnamefont {G.}~\bibnamefont {Cardoso}},
  \bibinfo {author} {\bibfnamefont {T.~H.}\ \bibnamefont {Hansson}},\ and\
  \bibinfo {author} {\bibfnamefont {Q.-D.}\ \bibnamefont {Jiang}},\ }\href
  {https://doi.org/10.1103/1tlw-g26r} {\bibfield  {journal} {\bibinfo
  {journal} {Phys. Rev. B}\ }\textbf {\bibinfo {volume} {113}},\ \bibinfo
  {pages} {045109} (\bibinfo {year} {2026})}\BibitemShut {NoStop}%
\bibitem [{\citenamefont {Wang}\ \emph {et~al.}(2026)\citenamefont {Wang},
  \citenamefont {Cardoso}, \citenamefont {Yang}, \citenamefont {Gong},
  \citenamefont {Zhang}, \citenamefont {Zhu}, \citenamefont {Zhang},
  \citenamefont {Pan}, \citenamefont {Cai}, \citenamefont {Chen}, \citenamefont
  {Jiang}, \citenamefont {Cheng}, \citenamefont {Wilczek},\ and\ \citenamefont
  {Zeng}}]{Wang2026}%
  \BibitemOpen
  \bibfield  {author} {\bibinfo {author} {\bibfnamefont {Z.}~\bibnamefont
  {Wang}}, \bibinfo {author} {\bibfnamefont {G.}~\bibnamefont {Cardoso}},
  \bibinfo {author} {\bibfnamefont {L.}~\bibnamefont {Yang}}, \bibinfo {author}
  {\bibfnamefont {X.}~\bibnamefont {Gong}}, \bibinfo {author} {\bibfnamefont
  {C.}~\bibnamefont {Zhang}}, \bibinfo {author} {\bibfnamefont
  {Y.}~\bibnamefont {Zhu}}, \bibinfo {author} {\bibfnamefont {D.}~\bibnamefont
  {Zhang}}, \bibinfo {author} {\bibfnamefont {N.}~\bibnamefont {Pan}}, \bibinfo
  {author} {\bibfnamefont {H.}~\bibnamefont {Cai}}, \bibinfo {author}
  {\bibfnamefont {Y.~P.}\ \bibnamefont {Chen}}, \bibinfo {author}
  {\bibfnamefont {Q.-D.}\ \bibnamefont {Jiang}}, \bibinfo {author}
  {\bibfnamefont {G.}~\bibnamefont {Cheng}}, \bibinfo {author} {\bibfnamefont
  {F.}~\bibnamefont {Wilczek}},\ and\ \bibinfo {author} {\bibfnamefont
  {C.}~\bibnamefont {Zeng}},\ }\href
  {https://doi.org/10.1038/s41586-026-11037-x} {\bibfield  {journal} {\bibinfo
  {journal} {Nature}\ }\textbf {\bibinfo {volume} {657}},\ \bibinfo {pages}
  {912} (\bibinfo {year} {2026})}\BibitemShut {NoStop}%
\bibitem [{\citenamefont {Appugliese}\ \emph {et~al.}(2022)\citenamefont
  {Appugliese}, \citenamefont {Enkner}, \citenamefont {Paravicini-Bagliani},
  \citenamefont {Beck}, \citenamefont {Reichl}, \citenamefont {Wegscheider},
  \citenamefont {Scalari}, \citenamefont {Ciuti},\ and\ \citenamefont
  {Faist}}]{Appugliese2022}%
  \BibitemOpen
  \bibfield  {author} {\bibinfo {author} {\bibfnamefont {F.}~\bibnamefont
  {Appugliese}}, \bibinfo {author} {\bibfnamefont {J.}~\bibnamefont {Enkner}},
  \bibinfo {author} {\bibfnamefont {G.~L.}\ \bibnamefont
  {Paravicini-Bagliani}}, \bibinfo {author} {\bibfnamefont {M.}~\bibnamefont
  {Beck}}, \bibinfo {author} {\bibfnamefont {C.}~\bibnamefont {Reichl}},
  \bibinfo {author} {\bibfnamefont {W.}~\bibnamefont {Wegscheider}}, \bibinfo
  {author} {\bibfnamefont {G.}~\bibnamefont {Scalari}}, \bibinfo {author}
  {\bibfnamefont {C.}~\bibnamefont {Ciuti}},\ and\ \bibinfo {author}
  {\bibfnamefont {J.}~\bibnamefont {Faist}},\ }\href
  {https://doi.org/10.1126/science.abl5818} {\bibfield  {journal} {\bibinfo
  {journal} {Science}\ }\textbf {\bibinfo {volume} {375}},\ \bibinfo {pages}
  {1030} (\bibinfo {year} {2022})}\BibitemShut {NoStop}%
\bibitem [{\citenamefont {Enkner}\ \emph {et~al.}(2025)\citenamefont {Enkner},
  \citenamefont {Graziotto}, \citenamefont {Bori{\c{c}}i}, \citenamefont
  {Appugliese}, \citenamefont {Reichl}, \citenamefont {Scalari}, \citenamefont
  {Regnault}, \citenamefont {Wegscheider}, \citenamefont {Ciuti},\ and\
  \citenamefont {Faist}}]{Enkner2025}%
  \BibitemOpen
  \bibfield  {author} {\bibinfo {author} {\bibfnamefont {J.}~\bibnamefont
  {Enkner}}, \bibinfo {author} {\bibfnamefont {L.}~\bibnamefont {Graziotto}},
  \bibinfo {author} {\bibfnamefont {D.}~\bibnamefont {Bori{\c{c}}i}}, \bibinfo
  {author} {\bibfnamefont {F.}~\bibnamefont {Appugliese}}, \bibinfo {author}
  {\bibfnamefont {C.}~\bibnamefont {Reichl}}, \bibinfo {author} {\bibfnamefont
  {G.}~\bibnamefont {Scalari}}, \bibinfo {author} {\bibfnamefont
  {N.}~\bibnamefont {Regnault}}, \bibinfo {author} {\bibfnamefont
  {W.}~\bibnamefont {Wegscheider}}, \bibinfo {author} {\bibfnamefont
  {C.}~\bibnamefont {Ciuti}},\ and\ \bibinfo {author} {\bibfnamefont
  {J.}~\bibnamefont {Faist}},\ }\href
  {https://doi.org/10.1038/s41586-025-08894-3} {\bibfield  {journal} {\bibinfo
  {journal} {Nature}\ }\textbf {\bibinfo {volume} {641}},\ \bibinfo {pages}
  {884} (\bibinfo {year} {2025})}\BibitemShut {NoStop}%
\bibitem [{\citenamefont {Jarc}\ \emph {et~al.}(2023)\citenamefont {Jarc},
  \citenamefont {Mathengattil}, \citenamefont {Montanaro}, \citenamefont
  {Giusti}, \citenamefont {Rigoni}, \citenamefont {Sergo}, \citenamefont
  {Fassioli}, \citenamefont {Winnerl}, \citenamefont {Dal~Zilio}, \citenamefont
  {Mihailovic}, \citenamefont {Prelov{\v{s}}ek}, \citenamefont {Eckstein},\
  and\ \citenamefont {Fausti}}]{Jarc2023}%
  \BibitemOpen
  \bibfield  {author} {\bibinfo {author} {\bibfnamefont {G.}~\bibnamefont
  {Jarc}}, \bibinfo {author} {\bibfnamefont {S.~Y.}\ \bibnamefont
  {Mathengattil}}, \bibinfo {author} {\bibfnamefont {A.}~\bibnamefont
  {Montanaro}}, \bibinfo {author} {\bibfnamefont {F.}~\bibnamefont {Giusti}},
  \bibinfo {author} {\bibfnamefont {E.~M.}\ \bibnamefont {Rigoni}}, \bibinfo
  {author} {\bibfnamefont {R.}~\bibnamefont {Sergo}}, \bibinfo {author}
  {\bibfnamefont {F.}~\bibnamefont {Fassioli}}, \bibinfo {author}
  {\bibfnamefont {S.}~\bibnamefont {Winnerl}}, \bibinfo {author} {\bibfnamefont
  {S.}~\bibnamefont {Dal~Zilio}}, \bibinfo {author} {\bibfnamefont
  {D.}~\bibnamefont {Mihailovic}}, \bibinfo {author} {\bibfnamefont
  {P.}~\bibnamefont {Prelov{\v{s}}ek}}, \bibinfo {author} {\bibfnamefont
  {M.}~\bibnamefont {Eckstein}},\ and\ \bibinfo {author} {\bibfnamefont
  {D.}~\bibnamefont {Fausti}},\ }\href
  {https://doi.org/10.1038/s41586-023-06596-2} {\bibfield  {journal} {\bibinfo
  {journal} {Nature}\ }\textbf {\bibinfo {volume} {622}},\ \bibinfo {pages}
  {487} (\bibinfo {year} {2023})}\BibitemShut {NoStop}%
\bibitem [{\citenamefont {Jiang}(2025)}]{Jiang2025}%
  \BibitemOpen
  \bibfield  {author} {\bibinfo {author} {\bibfnamefont {Q.-D.}\ \bibnamefont
  {Jiang}},\ }\href {https://doi.org/10.1038/s44310-025-00091-4} {\bibfield
  {journal} {\bibinfo  {journal} {npj Nanophoton.}\ }\textbf {\bibinfo {volume}
  {2}},\ \bibinfo {pages} {46} (\bibinfo {year} {2025})}\BibitemShut {NoStop}%
\bibitem [{\citenamefont {Cardoso}\ \emph {et~al.}(2026)\citenamefont
  {Cardoso}, \citenamefont {Yang}, \citenamefont {Hansson},\ and\ \citenamefont
  {Jiang}}]{Cardoso2026}%
  \BibitemOpen
  \bibfield  {author} {\bibinfo {author} {\bibfnamefont {G.}~\bibnamefont
  {Cardoso}}, \bibinfo {author} {\bibfnamefont {L.}~\bibnamefont {Yang}},
  \bibinfo {author} {\bibfnamefont {T.~H.}\ \bibnamefont {Hansson}},\ and\
  \bibinfo {author} {\bibfnamefont {Q.-D.}\ \bibnamefont {Jiang}},\ }\href
  {https://doi.org/10.1103/l7vh-p76b} {\bibfield  {journal} {\bibinfo
  {journal} {Phys. Rev. B}\ }\textbf {\bibinfo {volume} {113}},\ \bibinfo
  {pages} {045108} (\bibinfo {year} {2026})}\BibitemShut {NoStop}%
\bibitem [{\citenamefont {Dicke}(1954)}]{Dicke1954}%
  \BibitemOpen
  \bibfield  {author} {\bibinfo {author} {\bibfnamefont {R.~H.}\ \bibnamefont
  {Dicke}},\ }\href {https://doi.org/10.1103/PhysRev.93.99} {\bibfield
  {journal} {\bibinfo  {journal} {Phys. Rev.}\ }\textbf {\bibinfo {volume}
  {93}},\ \bibinfo {pages} {99} (\bibinfo {year} {1954})}\BibitemShut {NoStop}%
\bibitem [{\citenamefont {Wang}\ and\ \citenamefont
  {Hioe}(1973)}]{WangHioe1973}%
  \BibitemOpen
  \bibfield  {author} {\bibinfo {author} {\bibfnamefont {Y.~K.}\ \bibnamefont
  {Wang}}\ and\ \bibinfo {author} {\bibfnamefont {F.~T.}\ \bibnamefont
  {Hioe}},\ }\href {https://doi.org/10.1103/PhysRevA.7.831} {\bibfield
  {journal} {\bibinfo  {journal} {Phys. Rev. A}\ }\textbf {\bibinfo {volume}
  {7}},\ \bibinfo {pages} {831} (\bibinfo {year} {1973})}\BibitemShut {NoStop}%
\bibitem [{\citenamefont {Sur}\ \emph {et~al.}(2026)\citenamefont {Sur},
  \citenamefont {Wang}, \citenamefont {Mahankali}, \citenamefont {Paschen},\
  and\ \citenamefont {Si}}]{Sur2026}%
  \BibitemOpen
  \bibfield  {author} {\bibinfo {author} {\bibfnamefont {S.}~\bibnamefont
  {Sur}}, \bibinfo {author} {\bibfnamefont {Y.}~\bibnamefont {Wang}}, \bibinfo
  {author} {\bibfnamefont {M.}~\bibnamefont {Mahankali}}, \bibinfo {author}
  {\bibfnamefont {S.}~\bibnamefont {Paschen}},\ and\ \bibinfo {author}
  {\bibfnamefont {Q.}~\bibnamefont {Si}},\ }\href
  {https://doi.org/10.1038/s41467-026-73112-1} {\bibfield  {journal} {\bibinfo
  {journal} {Nat. Commun.}\ }\textbf {\bibinfo {volume} {17}},\ \bibinfo
  {pages} {4404} (\bibinfo {year} {2026})}\BibitemShut {NoStop}%
\bibitem [{\citenamefont {Zwettler}\ \emph {et~al.}(2025)\citenamefont
  {Zwettler}, \citenamefont {Marijanovic}, \citenamefont {B{\"u}hler},
  \citenamefont {Chattopadhyay}, \citenamefont {Del~Pace}, \citenamefont
  {Skolc}, \citenamefont {Helson}, \citenamefont {Uchino}, \citenamefont
  {Demler},\ and\ \citenamefont {Brantut}}]{Zwettler2025}%
  \BibitemOpen
  \bibfield  {author} {\bibinfo {author} {\bibfnamefont {T.}~\bibnamefont
  {Zwettler}}, \bibinfo {author} {\bibfnamefont {F.}~\bibnamefont
  {Marijanovic}}, \bibinfo {author} {\bibfnamefont {T.}~\bibnamefont
  {B{\"u}hler}}, \bibinfo {author} {\bibfnamefont {S.}~\bibnamefont
  {Chattopadhyay}}, \bibinfo {author} {\bibfnamefont {G.}~\bibnamefont
  {Del~Pace}}, \bibinfo {author} {\bibfnamefont {L.}~\bibnamefont {Skolc}},
  \bibinfo {author} {\bibfnamefont {V.}~\bibnamefont {Helson}}, \bibinfo
  {author} {\bibfnamefont {S.}~\bibnamefont {Uchino}}, \bibinfo {author}
  {\bibfnamefont {E.}~\bibnamefont {Demler}},\ and\ \bibinfo {author}
  {\bibfnamefont {J.-P.}\ \bibnamefont {Brantut}},\ }\href
  {https://doi.org/10.1038/s41467-025-67184-8} {\bibfield  {journal} {\bibinfo
  {journal} {Nat. Commun.}\ }\textbf {\bibinfo {volume} {17}},\ \bibinfo
  {pages} {496} (\bibinfo {year} {2025})}\BibitemShut {NoStop}%
\bibitem [{\citenamefont {Bethe}(1931)}]{Bethe1931}%
  \BibitemOpen
  \bibfield  {author} {\bibinfo {author} {\bibfnamefont {H.}~\bibnamefont
  {Bethe}},\ }\href {https://doi.org/https://doi.org/10.1007/BF01341708}
  {\bibfield  {journal} {\bibinfo  {journal} {Zeitschrift für Physik}\
  }\textbf {\bibinfo {volume} {71}},\ \bibinfo {pages} {205} (\bibinfo {year}
  {1931})}\BibitemShut {NoStop}%
\bibitem [{\citenamefont {Griffiths}(1964)}]{Griffiths1964}%
  \BibitemOpen
  \bibfield  {author} {\bibinfo {author} {\bibfnamefont {R.~B.}\ \bibnamefont
  {Griffiths}},\ }\href {https://doi.org/10.1103/PhysRev.133.A768} {\bibfield
  {journal} {\bibinfo  {journal} {Phys. Rev.}\ }\textbf {\bibinfo {volume}
  {133}},\ \bibinfo {pages} {A768} (\bibinfo {year} {1964})}\BibitemShut
  {NoStop}%
\bibitem [{\citenamefont {Yang}\ and\ \citenamefont
  {Yang}(1966)}]{YangYang1966}%
  \BibitemOpen
  \bibfield  {author} {\bibinfo {author} {\bibfnamefont {C.~N.}\ \bibnamefont
  {Yang}}\ and\ \bibinfo {author} {\bibfnamefont {C.~P.}\ \bibnamefont
  {Yang}},\ }\href {https://doi.org/10.1103/PhysRev.150.321} {\bibfield
  {journal} {\bibinfo  {journal} {Phys. Rev.}\ }\textbf {\bibinfo {volume}
  {150}},\ \bibinfo {pages} {321} (\bibinfo {year} {1966})}\BibitemShut
  {NoStop}%
\bibitem [{\citenamefont {Takahashi}(1971)}]{Takahashi1971}%
  \BibitemOpen
  \bibfield  {author} {\bibinfo {author} {\bibfnamefont {M.}~\bibnamefont
  {Takahashi}},\ }\href {https://doi.org/10.1143/PTP.46.401} {\bibfield
  {journal} {\bibinfo  {journal} {Prog. Theor. Phys}\ }\textbf {\bibinfo
  {volume} {46}},\ \bibinfo {pages} {401} (\bibinfo {year} {1971})}\BibitemShut
  {NoStop}%
\bibitem [{\citenamefont {Giamarchi}(2003)}]{Giamarchi2004}%
  \BibitemOpen
  \bibfield  {author} {\bibinfo {author} {\bibfnamefont {T.}~\bibnamefont
  {Giamarchi}},\ }\href
  {https://doi.org/10.1093/acprof:oso/9780198525004.001.0001} {\emph {\bibinfo
  {title} {Quantum Physics in One Dimension}}}\ (\bibinfo  {publisher} {Oxford
  University Press},\ \bibinfo {year} {2003})\BibitemShut {NoStop}%
\bibitem [{\citenamefont {Hammar}\ \emph {et~al.}(1999)\citenamefont {Hammar},
  \citenamefont {Stone}, \citenamefont {Reich}, \citenamefont {Broholm},
  \citenamefont {Gibson}, \citenamefont {Turnbull}, \citenamefont {Landee},\
  and\ \citenamefont {Oshikawa}}]{Hammar1999}%
  \BibitemOpen
  \bibfield  {author} {\bibinfo {author} {\bibfnamefont {P.~R.}\ \bibnamefont
  {Hammar}}, \bibinfo {author} {\bibfnamefont {M.~B.}\ \bibnamefont {Stone}},
  \bibinfo {author} {\bibfnamefont {D.~H.}\ \bibnamefont {Reich}}, \bibinfo
  {author} {\bibfnamefont {C.}~\bibnamefont {Broholm}}, \bibinfo {author}
  {\bibfnamefont {P.~J.}\ \bibnamefont {Gibson}}, \bibinfo {author}
  {\bibfnamefont {M.~M.}\ \bibnamefont {Turnbull}}, \bibinfo {author}
  {\bibfnamefont {C.~P.}\ \bibnamefont {Landee}},\ and\ \bibinfo {author}
  {\bibfnamefont {M.}~\bibnamefont {Oshikawa}},\ }\href
  {https://doi.org/10.1103/PhysRevB.59.1008} {\bibfield  {journal} {\bibinfo
  {journal} {Phys. Rev. B}\ }\textbf {\bibinfo {volume} {59}},\ \bibinfo
  {pages} {1008} (\bibinfo {year} {1999})}\BibitemShut {NoStop}%
\bibitem [{\citenamefont {Stone}\ \emph {et~al.}(2003)\citenamefont {Stone},
  \citenamefont {Reich}, \citenamefont {Broholm}, \citenamefont {Lefmann},
  \citenamefont {Rischel}, \citenamefont {Landee},\ and\ \citenamefont
  {Turnbull}}]{Stone2003}%
  \BibitemOpen
  \bibfield  {author} {\bibinfo {author} {\bibfnamefont {M.~B.}\ \bibnamefont
  {Stone}}, \bibinfo {author} {\bibfnamefont {D.~H.}\ \bibnamefont {Reich}},
  \bibinfo {author} {\bibfnamefont {C.}~\bibnamefont {Broholm}}, \bibinfo
  {author} {\bibfnamefont {K.}~\bibnamefont {Lefmann}}, \bibinfo {author}
  {\bibfnamefont {C.}~\bibnamefont {Rischel}}, \bibinfo {author} {\bibfnamefont
  {C.~P.}\ \bibnamefont {Landee}},\ and\ \bibinfo {author} {\bibfnamefont
  {M.~M.}\ \bibnamefont {Turnbull}},\ }\href
  {https://doi.org/10.1103/PhysRevLett.91.037205} {\bibfield  {journal}
  {\bibinfo  {journal} {Phys. Rev. Lett.}\ }\textbf {\bibinfo {volume} {91}},\
  \bibinfo {pages} {037205} (\bibinfo {year} {2003})}\BibitemShut {NoStop}%
\bibitem [{\citenamefont {Kono}\ \emph {et~al.}(2015)\citenamefont {Kono},
  \citenamefont {Sakakibara}, \citenamefont {Aoyama}, \citenamefont {Hotta},
  \citenamefont {Turnbull}, \citenamefont {Landee},\ and\ \citenamefont
  {Takano}}]{Kono2015}%
  \BibitemOpen
  \bibfield  {author} {\bibinfo {author} {\bibfnamefont {Y.}~\bibnamefont
  {Kono}}, \bibinfo {author} {\bibfnamefont {T.}~\bibnamefont {Sakakibara}},
  \bibinfo {author} {\bibfnamefont {C.~P.}\ \bibnamefont {Aoyama}}, \bibinfo
  {author} {\bibfnamefont {C.}~\bibnamefont {Hotta}}, \bibinfo {author}
  {\bibfnamefont {M.~M.}\ \bibnamefont {Turnbull}}, \bibinfo {author}
  {\bibfnamefont {C.~P.}\ \bibnamefont {Landee}},\ and\ \bibinfo {author}
  {\bibfnamefont {Y.}~\bibnamefont {Takano}},\ }\href
  {https://doi.org/10.1103/PhysRevLett.114.037202} {\bibfield  {journal}
  {\bibinfo  {journal} {Phys. Rev. Lett.}\ }\textbf {\bibinfo {volume} {114}},\
  \bibinfo {pages} {037202} (\bibinfo {year} {2015})}\BibitemShut {NoStop}%
\bibitem [{\citenamefont {Breunig}\ \emph {et~al.}(2017)\citenamefont
  {Breunig}, \citenamefont {Garst}, \citenamefont {Klümper}, \citenamefont
  {Rohrkamp}, \citenamefont {Turnbull},\ and\ \citenamefont
  {Lorenz}}]{Breunig2017}%
  \BibitemOpen
  \bibfield  {author} {\bibinfo {author} {\bibfnamefont {O.}~\bibnamefont
  {Breunig}}, \bibinfo {author} {\bibfnamefont {M.}~\bibnamefont {Garst}},
  \bibinfo {author} {\bibfnamefont {A.}~\bibnamefont {Klümper}}, \bibinfo
  {author} {\bibfnamefont {J.}~\bibnamefont {Rohrkamp}}, \bibinfo {author}
  {\bibfnamefont {M.~M.}\ \bibnamefont {Turnbull}},\ and\ \bibinfo {author}
  {\bibfnamefont {T.}~\bibnamefont {Lorenz}},\ }\href
  {https://doi.org/10.1126/sciadv.aao3773} {\bibfield  {journal} {\bibinfo
  {journal} {Science Advances}\ }\textbf {\bibinfo {volume} {3}},\ \bibinfo
  {pages} {eaao3773} (\bibinfo {year} {2017})}\BibitemShut {NoStop}%
\bibitem [{\citenamefont {Lake}\ \emph {et~al.}(2013)\citenamefont {Lake},
  \citenamefont {Tennant}, \citenamefont {Caux}, \citenamefont {Barthel},
  \citenamefont {Schollw{\"o}ck}, \citenamefont {Nagler},\ and\ \citenamefont
  {Frost}}]{Lake2013}%
  \BibitemOpen
  \bibfield  {author} {\bibinfo {author} {\bibfnamefont {B.}~\bibnamefont
  {Lake}}, \bibinfo {author} {\bibfnamefont {D.~A.}\ \bibnamefont {Tennant}},
  \bibinfo {author} {\bibfnamefont {J.-S.}\ \bibnamefont {Caux}}, \bibinfo
  {author} {\bibfnamefont {T.}~\bibnamefont {Barthel}}, \bibinfo {author}
  {\bibfnamefont {U.}~\bibnamefont {Schollw{\"o}ck}}, \bibinfo {author}
  {\bibfnamefont {S.~E.}\ \bibnamefont {Nagler}},\ and\ \bibinfo {author}
  {\bibfnamefont {C.~D.}\ \bibnamefont {Frost}},\ }\href
  {https://doi.org/10.1103/PhysRevLett.111.137205} {\bibfield  {journal}
  {\bibinfo  {journal} {Phys. Rev. Lett.}\ }\textbf {\bibinfo {volume} {111}},\
  \bibinfo {pages} {137205} (\bibinfo {year} {2013})}\BibitemShut {NoStop}%
\bibitem [{\citenamefont {Alcalde}\ \emph {et~al.}(2011)\citenamefont
  {Alcalde}, \citenamefont {Stephany},\ and\ \citenamefont
  {Svaiter}}]{AparicioAlcalde2011}%
  \BibitemOpen
  \bibfield  {author} {\bibinfo {author} {\bibfnamefont {M.~A.}\ \bibnamefont
  {Alcalde}}, \bibinfo {author} {\bibfnamefont {J.}~\bibnamefont {Stephany}},\
  and\ \bibinfo {author} {\bibfnamefont {N.~F.}\ \bibnamefont {Svaiter}},\
  }\href {https://doi.org/10.1088/1751-8113/44/50/505301} {\bibfield  {journal}
  {\bibinfo  {journal} {J. Phys. A: Math. Theor.}\ }\textbf {\bibinfo {volume}
  {44}},\ \bibinfo {pages} {505301} (\bibinfo {year} {2011})}\BibitemShut
  {NoStop}%
\bibitem [{\citenamefont {Johnson}\ and\ \citenamefont
  {Fowler}(1986)}]{Johnson1986}%
  \BibitemOpen
  \bibfield  {author} {\bibinfo {author} {\bibfnamefont {M.~D.}\ \bibnamefont
  {Johnson}}\ and\ \bibinfo {author} {\bibfnamefont {M.}~\bibnamefont
  {Fowler}},\ }\href {https://doi.org/10.1103/PhysRevB.34.1728} {\bibfield
  {journal} {\bibinfo  {journal} {Phys. Rev. B}\ }\textbf {\bibinfo {volume}
  {34}},\ \bibinfo {pages} {1728} (\bibinfo {year} {1986})}\BibitemShut
  {NoStop}%
\bibitem [{\citenamefont {Landau}\ and\ \citenamefont
  {Lifshitz}(1980)}]{Landau1980}%
  \BibitemOpen
  \bibfield  {author} {\bibinfo {author} {\bibfnamefont {L.~D.}\ \bibnamefont
  {Landau}}\ and\ \bibinfo {author} {\bibfnamefont {E.~M.}\ \bibnamefont
  {Lifshitz}},\ }\href {https://doi.org/10.1016/B978-0-08-057046-4.50019-1}
  {\emph {\bibinfo {title} {Statistical Physics}}},\ \bibinfo {edition} {3rd}\
  ed.,\ Vol.~\bibinfo {volume} {5}\ (\bibinfo  {publisher}
  {Butterworth-Heinemann},\ \bibinfo {year} {1980})\BibitemShut {NoStop}%
\bibitem [{\citenamefont {He}\ \emph {et~al.}(2017)\citenamefont {He},
  \citenamefont {Jiang}, \citenamefont {Yu}, \citenamefont {Lin},\ and\
  \citenamefont {Guan}}]{He2017}%
  \BibitemOpen
  \bibfield  {author} {\bibinfo {author} {\bibfnamefont {F.}~\bibnamefont
  {He}}, \bibinfo {author} {\bibfnamefont {Y.}~\bibnamefont {Jiang}}, \bibinfo
  {author} {\bibfnamefont {Y.-C.}\ \bibnamefont {Yu}}, \bibinfo {author}
  {\bibfnamefont {H.-Q.}\ \bibnamefont {Lin}},\ and\ \bibinfo {author}
  {\bibfnamefont {X.-W.}\ \bibnamefont {Guan}},\ }\href
  {https://doi.org/10.1103/PhysRevB.96.220401} {\bibfield  {journal} {\bibinfo
  {journal} {Phys. Rev. B}\ }\textbf {\bibinfo {volume} {96}},\ \bibinfo
  {pages} {220401(R)} (\bibinfo {year} {2017})}\BibitemShut {NoStop}%
\end{thebibliography}%

\clearpage
\onecolumngrid
\begin{center}
 {\large\bfseries End Matter}
\end{center}
\twocolumngrid
\setcounter{equation}{0}
\setcounter{figure}{0}
\setcounter{table}{0}
\renewcommand{\theequation}{E.\arabic{equation}}
\renewcommand{\thefigure}{E.\arabic{figure}}
\renewcommand{\thetable}{E.\arabic{table}}
\renewcommand{\theHequation}{endmatter.\arabic{equation}}
\renewcommand{\theHfigure}{endmatter.\arabic{figure}}
\renewcommand{\theHtable}{endmatter.\arabic{table}}

\textit{Free energy in the static saddle-point approximation}.\textemdash\
For $Z=\tr_{\mathrm{ph},s}e^{-\beta H}$ and $\beta=1/T$, the finite-system free-energy density is $f_{\mathrm{eq},N}=-T\ln Z/N$.
Inserting photon coherent states, with $a\ket\alpha=\alpha\ket\alpha$ and $\int d^2\alpha\,\ket\alpha\bra\alpha/\pi=1$, between imaginary-time slices gives
\begin{equation}
 \begin{aligned}
 Z&=\int\mathcal D[\alpha^*,\alpha]\,e^{-S_{\mathrm{eff}}}, \;\; S_{\mathrm{eff}} = S_{\mathrm{ph}} - \ln \mathcal Z_s[\phi]\\
 S_{\mathrm{ph}}&=\int_0^\beta d\tau\,
 \left[\alpha^*\partial_\tau\alpha+\omega|\alpha|^2\right],\\
 \mathcal Z_s[\phi]&=\tr_s\mathcal T_\tau
 \exp\!\left\{-\int_0^\beta d\tau\,
 [H_s(h)-\phi(\tau)M_z]\right\},
 \end{aligned}
 \label{eq:photon_path_integral}
\end{equation}
where $\phi(\tau)=g[\alpha(\tau)+\alpha^*(\tau)]/\sqrt N$ and $M_z=\sum_jS_j^z$.
Here $\mathcal T_\tau$ denotes imaginary-time ordering, $\tr_s$ denotes the trace over spins, and the measure integrates over periodic paths satisfying $\alpha(\beta)=\alpha(0)$.

Rescaling $\alpha=\sqrt N z$ makes the extensivity explicit.
We write the static saddle point as $z_0=x+iy$ and neglect photon fluctuations, 
as in path-integral treatments of Dicke-type models~\cite{AparicioAlcalde2011}.
With $\phi=2gx$,
\begin{equation}
 \frac{S_{\mathrm{eff}}^{\mathrm{stat}}}{\beta N}
 =\omega(x^2+y^2)+f_{\text{lm},N}(T,2gx,h),
 \label{eq:static_action}
\end{equation}
where the finite-chain spin free-energy density is
\begin{equation}
 f_{\mathrm{lm},N}(T,\phi,h)
 =-\frac{T}{N}\ln\tr_s e^{-\beta[H_s(h)-\phi M_z]}.
\end{equation}
Since only the cavity term depends on $y$, minimization gives $y=0$ for $\omega>0$.
Taking $f_\mathrm{lm}=\lim_{N\to\infty}f_{\mathrm{lm},N}$ and minimizing over $\phi=2gx$ yields Eq.~\eqref{eq:freeenergy}.
$\alpha_0=\sqrt N\phi_0/(2g)$ gives the photon density $\phi_0^2/(4g^2)$.

\textit{Coexistence and scaling near saturation}.\textemdash\
The polarized SR minimum has $\phi_0^2=q^2-h^2$.
At the first-order phase transition boundary, free energy density of SR minimum is equal to $e_{\mathrm{s}}(h)$, yielding Eq.~\eqref{eq:coexistence}.
Denote the energy-density difference between the ground state of $H_{\mathrm{s}}$ and the fully polarized state as $D(h)$, 
\begin{equation}
    D(h)=\frac{1}{4}-\frac{h}{2}-e_{\mathrm{s}}(h),
\end{equation}
we obtain
\begin{equation}
 q_c=h+2D(h)+2\sqrt{D(h)[h+D(h)]}.
 \label{eq:qc_exact}
\end{equation}

Near the saturation transition, the ground state of $H_{\mathrm{lm}}$ is described by a dilute magnon gas excited from the fully polarized state \cite{Johnson1986}.
For momentum $p$ measured from the magnon minimum, the one-magnon dispersion is
$\varepsilon(p)=h-2+1-\cos p=-\delta h+p^2/2+O(p^4)$.
In the dilute limit, the leading energy density is that of impenetrable magnons, equivalently that of spinless fermions filling the negative-energy modes.
Thus, with $p_F=\sqrt{2\delta h}$,
\begin{equation}
 \begin{aligned}
 D(h)&=\int_{-p_F}^{p_F}\frac{dp}{2\pi}
 \left(\delta h-\frac{p^2}{2}\right)+o(\delta h^{2})\\
 &=A\delta h^{3/2}+o(\delta h^{2}),
 \end{aligned}
\end{equation}
where $A=2\sqrt2/(3\pi)$.
Eq.~\eqref{eq:qc_exact} then gives $q_c-2\sim C\delta h^{3/4},\ \Delta\phi\sim2\sqrt C\,\delta h^{3/8}$ with $C=2\sqrt{2A}$.

Exactly at $h=2$, the accessible effective fields satisfy $b\geq2$, and the polarized expression gives
\begin{equation}
 \mathcal F(\phi)-\mathcal F(0)
 =\left(\frac{1}{4q}-\frac18\right)\phi^2
 +\frac{\phi^4}{128}+O(\phi^6).
 \label{eq:zero_endpoint_expansion}
\end{equation}
The positive quartic coefficient at $q=2$ excludes an ordinary $r=u=0$ tricritical description at this point.
For $h<2$, the nonanalyticity at $b=2$ lies at $|\phi|=\sqrt{4-h^2}$, which tends to zero as $h\to2^-$.
The small-$\phi$ expansion about the gapless normal state is therefore nonuniform in this limit and cannot describe the coexistence minimum beyond saturation.

\textit{Low-$T$ regime that exhibits 3/8 scaling}.\textemdash\
We show that in the temperature window where $T\ll\delta h\ll1$ with $\delta h=2-h>0$, the first-order jump is dominated by its zero $T$ behavior. 
For the SR minimum, because $\sqrt{\Delta \phi^2+h^2}>2$, the excitations have an energy gap that is much greater than $T$. 
Therefore, the free energy density can be approximated as that of the zero-$T$ case.
Similar to the previous section, we define the finite $T$ counterpart of $D(h)$ as 
\begin{equation}
    D_T(h)=\frac{1}{4}-\frac{h}{2}-f_{\mathrm{s}}(T,h).
\end{equation}
Following the same procedure as in the previous section, we have
\begin{equation}
    q_c=h+2D_T(h)+2\sqrt{D_T(h)[h+D_T(h)]}.
    \label{eq:finite_Dt}
\end{equation}
\par
Using the dilute magnon gas description, we have
\begin{equation}
    D_T(h)=T\int_{-\infty}^{\infty}\frac{dp}{2\pi}\ln\!\left[1+e^{(\delta h-p^2/2)/T}\right].
\end{equation}
Therefore, we can estimate the difference between $D_T(h)$ and $D(h)$ following Sommerfeld expansion \cite{Landau1980},
\begin{equation}
 D_T(h)-D(h)=\frac{\pi T^2}{6\sqrt{2\delta h}},
 \label{eq:lowT_expansion1}
\end{equation}
and relative difference
\begin{equation}
 \frac{D_T(h)}{D(h)}=1+\frac{\pi^2}{8}
 \left(\frac{T}{\delta h}\right)^2.
 \label{eq:lowT_expansion2}
\end{equation}
Substituting $D_T(h)$ into Eq.~\eqref{eq:finite_Dt}, we obtain Eq.~\eqref{eq:lowT_jump_recovery}.
\par
\textit{Finite-$T$ Landau coefficients}.\textemdash\
For $T>0$ and $h>0$, expanding the spin free energy and including the cavity term gives the coefficients in Eq.~\eqref{eq:thermal_landau_main},
\begin{equation}
 \begin{aligned}
 r&=\frac{1}{2q}-\chi_\perp,\\
 u&=\frac{\chi_\perp-\chi_\parallel}{2h^2},\\
 v&=\frac{-3\chi_\perp+3\chi_\parallel-h\partial_h\chi_\parallel}{8h^4},
 \end{aligned}
 \label{eq:landau_coefficients}
\end{equation}
where $m=-\partial_hf_{\mathrm{s}}(T,h)$, $\chi_\perp=m/h$, and $\chi_\parallel=\partial_hm$.
At $h=0$, the free energy is even and analytic in the field, so a direct expansion gives
$
 r=\frac{1}{2q}+\partial_\phi^2f_{\mathrm{lm}}(T,\sqrt{\phi^2})|_{\phi=0},\ 
 u=\frac16\partial_\phi^4f_{\mathrm{lm}}(T,\sqrt{\phi^2})|_{\phi=0},\ 
 v=\frac1{120}\partial_\phi^6f_{\mathrm{lm}}(T,\sqrt{\phi^2})|_{\phi=0}.
$
The zero-field conditions $r=u=0$ give
\begin{equation}
 \partial_\phi^4 f_{\mathrm{s}}(T_{\mathrm{tc}}^{(0)},0)=0,\qquad 
 q_{\mathrm{tc}}^{(0)}=-\frac{1}{2\partial_\phi^2f_{\mathrm{lm}}(T,\sqrt{\phi^2})|_{\phi=0}}.
 \label{eq:zero_field_endpoint}
\end{equation}

On the first-order phase transition boundary, setting $X=\Delta\phi^2>0$, the coexistence relation requires
\begin{equation}
 r+uX+vX^2=0,\qquad
 \frac r2+\frac{uX}{4}+\frac{vX^2}{6}=0.
\end{equation}
Eliminating $r$ gives Eq.~\eqref{eq:finiteT_coexistence}.
The intervening maximum at $\phi^2=-u/(4v)$ has height $(-u)^3/(96v^2)$.

\begin{table}[t]
 \centering
 \begin{tabular}{cccc}
 \hline\hline
 $N$ & $T=0.05$ & $T=0.5$ & $T=1.0$ \\
 \hline
 12 & 2.093800 & 2.114207 & 0.836949 \\
 14 & 2.091853 & 2.114183 & 0.836931 \\
 16 & 2.090953 & 2.114180 & 0.836930 \\
 18 & 2.090506 & 2.114180 & 0.836930 \\
 20 & 2.090290 & 2.114180 & 0.836930 \\
 \hline\hline
 \end{tabular}
 \caption{Finite-chain endpoint estimates $h_{\mathrm{tc},N}$, with $J=1$.
 Six decimal places are reported to display the finite-size shifts. Accuracy is assessed relative to $\Delta h=5\times10^{-4}$, not the number of displayed digits.}
 \label{tab:convergence}
\end{table}

Because $\partial_h\chi_\perp=(\chi_\parallel-\chi_\perp)/h$ vanishes at a nonzero-field endpoint,
differentiating $u$ and evaluating $v$ there gives
\begin{equation}
 \begin{aligned}
 v_{\mathrm{tc}}&=\left.-\frac{\partial_h\chi_\parallel}{8h^3}\right|_{h=h_{\mathrm{tc}}},\\
 u_h&=\partial_hu|_{h=h_{\mathrm{tc}}}
 =4h_{\mathrm{tc}}v_{\mathrm{tc}}.
 \end{aligned}
 \label{eq:endpoint_slope}
\end{equation}
Together with Eq.~\eqref{eq:finiteT_coexistence}, these relations give the leading fixed-$T$ result $\Delta\phi^2\simeq3h_{\mathrm{tc}}\delta h_T$ used in the main-text scaling argument.

\textit{Numerical determination of thermal endpoints}.\textemdash\
At fixed $T>0$, we locate a nonzero-field endpoint from the zero of
\begin{equation}
 R(T,h)\equiv h\chi_\parallel(T,h)-m(T,h),
 \qquad h>0,
 \label{eq:endpoint_residual}
\end{equation}
and obtain $q_{\mathrm{tc}}=h_{\mathrm{tc}}/[2m(T,h_{\mathrm{tc}})]$.
At $h=0$, symmetry makes $R$ vanish identically, so we instead use Eq.~\eqref{eq:zero_field_endpoint}.
The numerical scanning grid has $\Delta h=5\times10^{-4}$ and $\Delta T=10^{-4}$.
We accept endpoint estimates when refinement shifts are negligible relative to $\Delta h$, so finer grids require tighter convergence.

At high $T$s, particularly when $h/T$ is small, thermally populated long strings can give power-law decaying contributions~\cite{He2017}, making convergence with respect to the TBA string cutoff costly at the resolution needed for $R$.
ED avoids the need for this string cutoff, and Table~\ref{tab:convergence} exhibits its performance on convergence at high $T$. 
At low $T$s, appreciable ED finite-size effects instead favor thermodynamic-limit TBA~\cite{Takahashi1971}.
The large value of $h/T$ near saturation suppresses long strings and permits efficient cutoff control~\cite{He2017}.
\end{document}